\documentclass[aps,prd,twocolumn,superscriptaddress,floatfix,preprintnumbers,showpacs,nofootinbib]{revtex4-1}
\usepackage{dcolumn}
\usepackage{amssymb}
\usepackage{amsmath}
\usepackage{epsfig}
\usepackage{color}
\usepackage{times}
\usepackage{bm}
\usepackage{latexsym}
\usepackage{ulem}
\usepackage{graphicx}
\usepackage{subfig}
\usepackage[colorlinks]{hyperref}

\newcommand{\ket}[1]{\vert #1 \rangle}
\newcommand{\bra}[1]{\langle #1 \vert}

\begin{document}

\preprint{ADP-12-08/T775}
\title{Neutralino-hadron scattering in the NMSSM}

\author{Sophie J. Underwood}
\affiliation{ARC Centre of Excellence for Particle Physics at the Terascale and CSSM, School of Chemistry and Physics, University of Adelaide 5005, Australia.}
\author{Joel Giedt}
\affiliation{Department of Physics, Applied Physics, and Astronomy, Rensselaer Polytechnic Institute, 110 Eighth Street, Troy, New York 12180-3590, USA.}
\author{Anthony W. Thomas}
\affiliation{ARC Centre of Excellence for Particle Physics at the Terascale and CSSM, School of Chemistry and Physics, University of Adelaide 5005, Australia.}
\author{Ross D. Young}
\affiliation{ARC Centre of Excellence for Particle Physics at the Terascale and CSSM, School of Chemistry and Physics, University of Adelaide 5005, Australia.}

\begin{abstract}

We provide a scan of the parameter space for neutralino-hadron scattering in the next-to-minimal supersymmetric standard model using an updated value for the strange quark sigma commutator. These results also take into account constraints from WMAP data on the relic density and new constraints from the Large Hadron Collider. We find that the resultant spin-independent cross sections are smaller in magnitude than those found in recent results obtained within the constrained minimal supersymmetric standard model, yet still great enough to feasibly allow for detection in the case of bino-like neutralinos.

\end{abstract}

\maketitle

It is now well established that a vast portion of our universe (almost a quarter) is comprised of a weakly interacting quantity known as dark matter. There is no satisfactory candidate within the standard model and thus its categorisation remains one of the great conundrums of modern physics. Currently, theories in which dark matter consists of slow-moving particles are most favoured, on the basis of extensive evidence from rotation curves, galaxy clusters, gravitational lensing and anisotropies in the cosmic microwave background radiation \cite{Bertone:2004pz}. In particular, Weakly Interacting Massive Particles (WIMPs) are the leading candidates. There are many detection experiments underway around the world that seek to establish the presence of such WIMPs, both directly (including CDMS, DAMA/LIBRA, DRIFT, EDELWEISS, LUX, PICASSO, SIMPLE, WArP, XENON, XMASS and ZEPLIN-III \cite{Akimov:2011zz,Armengaud:2011cy,Felizardo:2010mi,Archambault:2009sm}) and indirectly (including AMANDA, ANTARES, Fermi-LAT, IceCube and PAMELA \cite{Morselli:2011zza,IceCube:2011aj,Bertin:2011zz}). It is vital to thoroughly explore the properties of dark matter, particularly its interactions with hadronic matter, within a variety of different theoretical models, since this will help to guide direct efforts to detect it.

Supersymmetric models have been quite popular within the past few decades - particularly the minimal supersymmetric standard model (MSSM). In these scenarios, dark matter takes the form of the lightest supersymmetric particle (LSP), usually a neutralino; it is favoured as a dark matter candidate for its stable and weakly-interacting nature \cite{Jungman:1995df}. Various authors in the past decade or so have used the MSSM and its variants to make predictions about its interaction with baryonic matter via the calculation of spin independent neutralino-hadron scattering cross sections. However, in the light of recent results from the LHC, the MSSM is becoming increasingly hard-pressed to serve as a complete description of physics beyond the standard model, because of the need to fine-tune its parameter space. The next-to-minimal supersymmetric standard model (NMSSM), on the other hand, is under less threat \cite{Djouadi:2008uw}, and is currently a model of considerable interest.

In this Letter, we calculate the spin-independent cross section for neutralino-nucleon scattering within the constrained NMSSM, taking into account the constraints from the initial running of the LHC, as well as lattice QCD determinations of the light quark sigma commutators. We incorporate the latest relic density constraints from WMAP results and find a drastic reduction of regions in the NMSSM parameter space for which neutralino dark matter is viable. We show that the spin-independent cross sections fall within two main regions; the first corresponding to bino-like neutralinos with cross sections on the edge of detection limits, the second to singlino-like neutralinos with cross-sections far too small to be detected in current experiments.

To begin, we briefly outline the relevant features of the NMSSM; for a comprehensive review, we recommend \cite{Ellwanger:2009dp}. The MSSM introduces two neutral Higgs doublets $H_u$ and $H_d$ to the standard model. The Higgs superfields contribute a Higgs mass term to the superpotential of the MSSM,
\begin{equation}
W_{MSSM} = W_Y + \mu \hat{H}_u \hat{H}_d,
\end{equation}
where $W_Y$ represents the Yukawa couplings for the SM fermions and $\hat H_u$, $\hat H_d$ are the Higgs chiral superfields. In order to avoid extreme fine-tuning, it is necessary that the $\mu$ term and the scale of SUSY breaking both lie at the electroweak scale. It is unknown why the two scales should fall so close to each other (and far below the GUT scale) when $\mu$ itself has little to do with SUSY breaking; this is generally considered to be a problem of naturalness.

Historically, the NMSSM was formulated as a convenient way of dealing with this ``$\mu$ problem.'' In the NMSSM, $\mu$ is replaced by a gauge singlet chiral superfield $\hat{S}$. An effective $\mu$ can thus be dynamically generated upon SUSY breaking, explaining the coincidence of scales. This results in an expanded superpotential in the NMSSM,
\begin{equation}
W_{NMSSM} = W_Y + \lambda \hat{S} \hat{H}_u\hat{H}_d + \frac{\kappa}{3}\hat{S}^3.
\end{equation}

The promotion of $\mu$ to a singlet field does have some consequences for the neutralino-hadron cross-section, since it results in a greater number of ways in which these particles can interact. Two extra Higgs fields are generated, such that the Higgs sector of the NMSSM consists of three neutral CP-even Higgs, two CP-odd Higgs and two charged Higgs. However, only the three CP-even Higgs are of relevance when formulating neutralino-hadron spin-independent cross sections.

The lightest neutralino itself may be described as a mixing of five neutral fields rather than the four of the MSSM, 
\begin{equation}
\chi = Z_{\chi^1} \tilde{B} + Z_{\chi^2} \tilde{W} + Z_{\chi^3} \tilde{H}_1 + Z_{\chi^4} \tilde{H}_2 + Z_{\chi^5} \tilde{S}.
\end{equation}
Here, $\tilde{B}$ is the bino (superpartner of the $U(1)$ gauge field), $\tilde{W}$ is the wino (superpartner of the $W$ gauge field), $\tilde{H}_1$ and $\tilde{H}_2$ are higgsinos (superpartners of the Higgs fields) and $\tilde{S}$ is the singlino (superpartner of the singlet field). The behaviour of the neutralino as it interacts with hadronic matter is strongly dependent on its exact composition. A predominantly bino-like neutralino (99$\%$ bino), for instance, will be shown to yield a high spin-independent cross-section, relative to a singlino-like neutralino, since singlinos do not couple to sfermions, quarks or gauge fields.

Before proceeding to the results, it may be helpful to outline the method used to calculate the spin-independent neutralino-hadron cross section, $\sigma_{SI}$. (We refrain from addressing the spin-dependent component of the neutralino-hadron cross section, since it is typically several orders of magnitude below experimental sensitivity \cite{Goodman:1984dc}.) 

\begin{figure}[!htbp]
\centering
\vspace{2mm}
\includegraphics[scale=0.35]{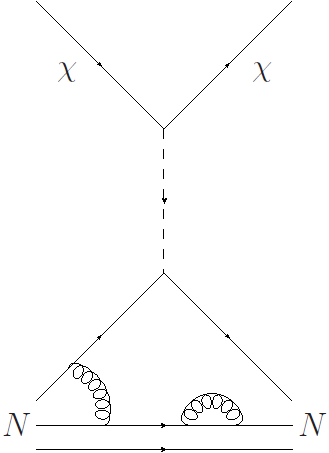}
\caption{A neutralino-nucleon collision via Higgs exchange.}
\label{collision}
\end{figure}

Fig.~\ref{collision} shows one example of an interaction between a neutralino (denoted by $\chi$) and a nucleon (neutralinos and nucleons may also interact via squark exchange). The cross section for this kind of interaction involves matrix elements of the form $\bra{A} \bar{q}q \ket{A}$. The following gives a succinct expression for the contribution to $\sigma_{SI}$ of each quark \cite{Ellis:2005mb,Ellis:2008hf,Hill:2011be}:
\begin{eqnarray*}
\bra{N \chi} \alpha_{3q} \bar{\chi}\chi \bar{q}q \ket{\chi N} & = & \alpha_{3q} \bra{N} \bar{q}q \ket{N} \\
& = & \alpha_{3q} m_N \frac{f_{Tq}}{m_q},
\end{eqnarray*}
where we have used $m_N f_{Tq} = \bra{N}m_q \bar{q}q\ket{N}$ for the sigma terms (the bottom half of Fig.~\ref{collision}) and $\alpha_{3q}$ encapsulates the relevant physics in terms of the amplitudes of each contributing neutralino-quark interaction (the top half of Fig.~\ref{collision}). Summing over light and heavy quarks gives a generic expression for $\sigma_{SI} = 4 \left(m^2_N/\pi\right) f^2$, where
\begin{equation}
\frac{f}{m_N} = \sum_{q=u,d,s} \frac{\alpha_{3q}f_{Tq}}{m_q} + \sum_{Q=c,b,t} \frac{\alpha_{3Q}f_{TQ}}{m_Q}.
\end{equation}
This may be simplified further by noting that $m_N = \bra{N} \theta^\mu_\mu \ket{N}$ for a system at rest, where the trace of the energy-momentum tensor is given by
\begin{equation}
\theta^\mu_\mu = \sum_q m_q \bar{q}q + \sum_Q m_Q \bar{Q}Q - \frac{7\alpha_s}{8\pi} G_{\mu\nu}G^{\mu\nu}.
\end{equation}
Taking the system to be at zero-momentum, such that $\theta^\mu_\mu = \theta^0_0$, this equation becomes
\begin{align}
m_N & = \bra{N}\sum_q m_q \bar{q}q\ket{N} + \bra{N}\sum_Q m_Q \bar{Q}Q\ket{N} \nonumber \\
& \hspace{4mm}- \bra{N}\frac{7\alpha_s}{8\pi}G_{\mu\nu}G^{\mu\nu}\ket{N}.
\end{align}
The extra term involving $G_{\mu\nu}G^{\mu\nu}$ can be conveniently eliminated by appealing to the Novikov-Shifman-Vainshtein \cite{Shifman:1978bx} relation, which tells us that $-(7\alpha_s/8\pi)\bra{N}G_{\mu\nu}G^{\mu\nu}\ket{N}$ may be written as $\left(7/2\right) \sum_Q m_Q \bra{N}\bar{Q}Q\ket{N}$,
\begin{eqnarray*}
m_N &=& \sum_q \bra{N} \bar{q}q \ket{N} + \tfrac{9}{2} \sum_Q m_Q \bra{N}\bar{Q}Q\ket{N} \\
&=& \sum_q m_N f_{Tq} + \tfrac{27}{2} m_Q \bra{N} \bar{Q}Q \ket{N},
\end{eqnarray*}
and thus $m_Q \bra{N} \bar{Q}Q \ket{N} = \tfrac{2}{27} m_N \Big(1 - \sum_q f_{Tq}\Big)$. Hence, the final expression for $f$ is
\begin{equation}
\begin{aligned}
\frac{f}{m_N} &= \sum_{q=u,d,s} \frac{\alpha_{3q} f_{Tq}}{m_q} + \tfrac{2}{27} f_{TQ}\sum_{Q=c,b,t} \frac{\alpha_{3Q}}{m_Q}, \\
f_{TQ} &= \Big(1 - \sum_{q=u,d,s} f_{Tq}\Big).
\end{aligned}
\end{equation}

The $\alpha_{3q}$ terms for the NMSSM themselves are derived by computing the amplitudes of the contributing Feynman diagrams. In addition, the $f_{Tq}$ terms have been derived from numerous studies in lattice QCD \cite{Giedt:2009mr,Young:2009zb,Toussaint:2009pz}, which are updates of early estimates \cite{Nelson:1987dg,Borasoy:1996bx}.

In order to compute the evolution of the coefficients $\alpha_{3q}$ from the GUT scale to the EW scale, we modified micrOMEGAs, a code for calculating general dark matter properties under supersymmetric physics, developed by Belanger {\it et al.} \cite{Belanger}. In order to reduce the parameter space, we chose a constrained version of the NMSSM in which the scalar and gaugino masses were taken to be universal at the GUT scale. Thus, the free parameters are the universal scalar mass $m_0$, universal gaugino mass $m_{1/2}$, singlino trilinear coupling $A_\kappa$, Higgs-singlino trilinear coupling $A_\lambda$ (later referred to simply as $A$ in this paper), tan$\beta$ (the ratio of vacuum expectation values of the neutral Higgses) and $\lambda$. In addition, the effective Higgs mass $\mu$ was taken to be positive. One of the constraints imposed by micrOMEGAs (namely, the computation of the muon anomalous moment \cite{Bennett:2006fi}) was relaxed.

\begin{figure*}
\centering
\subfloat[tan$\beta = 50$, $\lambda = 0.01$, $A_\kappa = -40$]{\label{relic50}\includegraphics[width=0.5\textwidth]{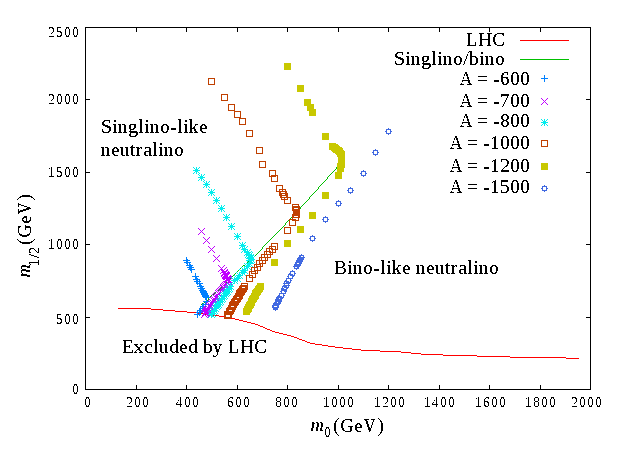}}
\subfloat[tan$\beta = 5$ or $10$, $\lambda = 0.1$, $A_\kappa = -40$]{\label{relic5and10}\includegraphics[width=0.5\textwidth]{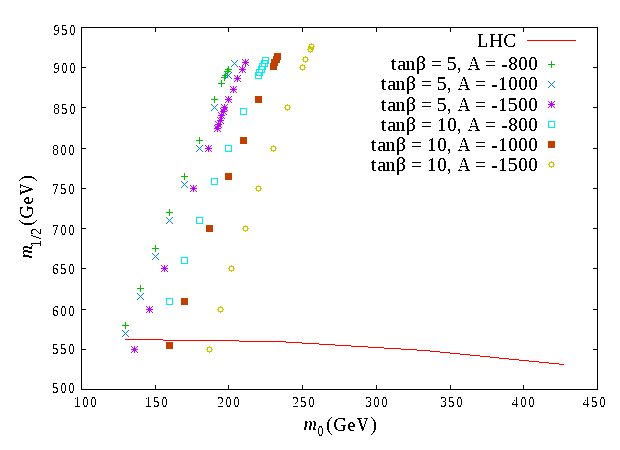}}
\caption{Regions in the space of universal spin-$1/2$ and spin-$0$ masses allowed by relic density constraints}
\end{figure*}

\begin{figure*}
\centering
\subfloat[tan$\beta = 50$, $\sigma_s = 22 \pm 6$, $\sigma_l = 47 \pm 9$]{\label{crosssections50}\includegraphics[width=0.5\textwidth]{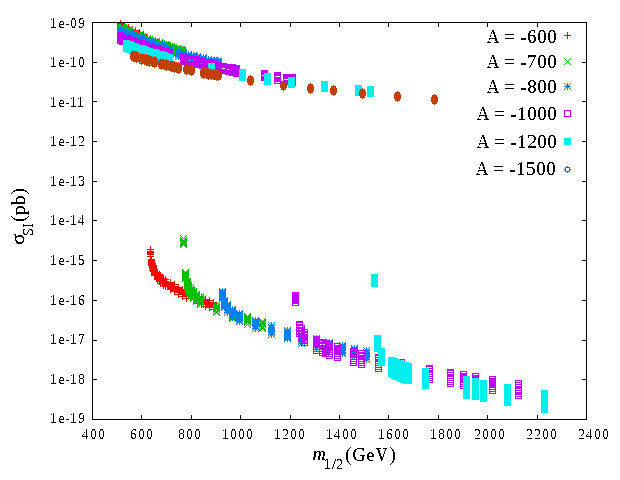}}
\subfloat[tan$\beta = 5$ or $10$, $\sigma_s = 22 \pm 6$, $\sigma_l = 47 \pm 9$]{\label{crosssections5and10}\includegraphics[width=0.5\textwidth]{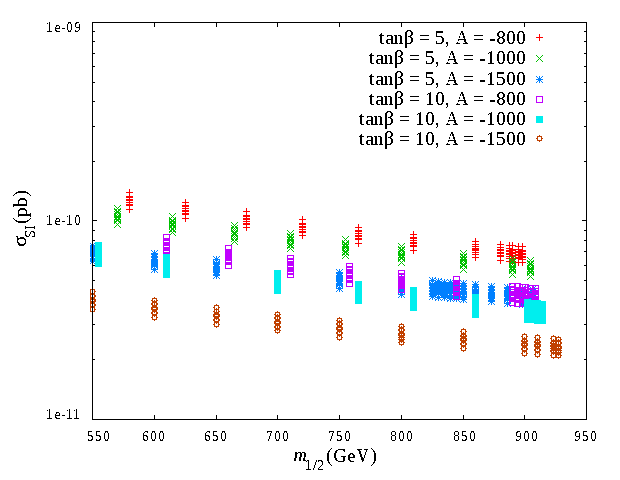}}
\caption{Cross sections for the spin-independent neutralino-neucleon cross section for the parameter sets illustrated in Fig.~\ref{relic50} and Fig.~\ref{relic5and10}}
\end{figure*}

Initially, the light quark sigma term $\sigma_l$ was taken to be 47 MeV, to correspond with a lattice determination of $\sigma_l = 47 \pm 9$ MeV, and the strangeness sigma term $\sigma_s$ was taken to be 50 MeV, in accordance with $\sigma_s = 50 \pm 8$ MeV from \cite{Giedt:2009mr}, which was obtained by averaging two different lattice results. However, recent findings tend to favour an even lower value \cite{Shanahan:2011su, Thomas:2012tg}. Thus, these scans were repeated with $\sigma_s = 22 \pm 6$ MeV and $\sigma_l = 47 \pm 9$ MeV, and the plots provided used these values. In addition, findings from WMAP and other observations have placed constraints on the relic density $\Omega$ to lie between 0.1053 and 0.1193 at 95$\%$ confidence level \cite{Herrmann:2010jp, Komatsu:2010fb}. This constraint places tight restrictions on the allowed parameter space, where for a fixed $A$, the allowed regions are reduced to thin strips or lines in the $(m_0,m_{1/2})$ plane. Finally, recent data from the CMS collaboration \cite{Chatrchyan:2011zy} was used to place a lower bound on $(m_0,m_{1/2})$, with similar results having been obtained from ATLAS \cite{Aad:2011ib}. Although this bound was originally formulated within the context of the MSSM, the spectrum of superpartners is quite similar in the NMSSM within this region of parameter space, so the bound still represents a very good approximation in the present case.

Sweeps of $m_0$ and $m_{1/2}$ were carried out for various values of $A$ at high tan$\beta$ (50), with $\lambda$ and $A_\kappa$ fixed at 0.01 and -40 respectively. Points that are allowed by both LHC and relic density constraints are plotted in Figure~\ref{relic50}. Noticeably, as $m_{1/2}$ is increased the character of the neutralino changes quite significantly. For lower $m_{1/2}$ along each strip, the content is predominantly bino, and cross-sections are of the order of $10^{-9}$ pb. However, as the points cross the bino-singlino line, as illustrated in Figure~\ref{relic50}, the neutralino is almost entirely singlino and the cross-section sharply drops by several orders of magnitude. For this reason, the region immediately to the right of the bino-singlino line is most favourable as an area of interest, since neutralinos in the singlino-like region have interaction cross sections that are far too small to allow for detection in current searches (see Figure~\ref{crosssections50}).

In addition, lower values of tan$\beta$ (5 and 10) also yield allowed regions, although the highest cross-sections are found for tan$\beta$ = 50 (see Figure~\ref{crosssections5and10}). However, in the low tan$\beta$ region we do not observe any singlino-like behaviour (as evidenced in Figure~\ref{relic5and10}), since a limit is quickly reached beyond which the LSP is no longer a neutralino. Very low values of tan$\beta$ proved to be unfavourable, because of the presence of a Landau pole.

\begin{table}[h!]
\caption{Example breakdown of quark flavour contributions at high tan$\beta$ with $\sigma_s = 22$ MeV, $\sigma_l = 47$ MeV.}
\centering
\begin{tabular}[t]{|c|c|c|c|}\hline
Model&$q$&$\alpha_{3q}/m_q$&$f^p_q / f_p$\\\hline\hline
tan$\beta$ = 50, $A = -575$&$u$&$-1.179 \times 10^{-9}$&0.0196\\
$m_0 = 436$, $m_{1/2} = 510$&$d$&$-1.090 \times 10^{-8}$&0.1820\\
$\lambda = 0.01$, $A_\kappa = -40$&$c$&$-1.179 \times 10^{-9}$&0.0538\\
$\sigma_{SI} = 8.678 \times 10^{-10}$ pb&$s$&$-1.090 \times 10^{-8}$&0.1700\\
$\sigma_l = 47$ MeV&$t$&$-1.174 \times 10^{-9}$&0.0536\\
$\sigma_s = 22$ MeV&$b$&$-1.142 \times 10^{-8}$&0.5213\\\hline
\end{tabular}
\label{table}
\end{table}

Table~\ref{table} shows that in the NMSSM at high tan$\beta$, the cross-section is dominated by the down-type quarks (particularly the bottom and strange quarks). This is similar to the finding in the CMSSM, although in the present case the bottom quark dominates by an even greater percentage. This is shown also for Table~\ref{table} with an updated $\sigma_s$ value; lowering $\sigma_s$ from 47 MeV to 22 MeV seems to have the effect of reducing the cross sections by typically 30$\%$.

From these results, it seems that the constrained NMSSM does indeed produce a number of viable dark matter candidates. $\sigma_{SI}$ for the neutralino-hadron collision in this scheme is very strongly dependent on the composition of the neutralino itself. In the region where the neutralino is predominantly bino-like, $\sigma_{SI}$ is comparable to the values found in the CMSSM and within the reach of direct detection experiments. On the other hand, there is a second region in which the neutralino is predominantly singlino-like, where $\sigma_{SI}$ is negligibly small. Furthermore, given the sharp drop in the $\sigma_{SI}$ for these singlino-like neutralinos, our results suggest a possible scenario in which a discovery at the LHC may be compatible with a null result in direct detection dark matter searches. Further investigation will be necessary for other variations of supersymmetric models as new data from the LHC is produced.

Sophie Underwood was supported by an ARC LF Postgraduate Researcher Scholarship (FL 0992247) and a Norman and Patricia Polglase Supplementary Scholarship. Joel Giedt was supported by the Dept.~of Energy, Office of Science, Office of High Energy Physics, Grant No.~DE-FG02-08ER41575. This work was also supported by the ARC Centre of Excellence in Particle Physics at the Terascale, by Australian Laureate Fellowship (FL0992247, Anthony Thomas) and by grant DP 110101265 (Ross Young).

\end{document}